\documentclass[12pt]{article}
\usepackage{times} 
\usepackage{fullpage}
\usepackage{amsmath,amscd,amssymb,graphics}
\usepackage{mathrsfs} 
\usepackage{color}

\newcommand{\abs}[1]{\lvert#1\rvert} 
\def\E{{\mathbb{E}}}

\def\R{{\mathbb R}}
\def\e{\varepsilon}

\begin{document}
\title{Intrinsic Dimensionality}

\author{Vladimir Pestov}
\date{\small Department of Mathematics and Statistics, University of Ottawa, Ottawa, Ontario, Canada}

\maketitle

\noindent{\bf Nature of the concept.}
Intrinsic dimensionality of data does not refer to a single well-defined parameter, but rather a diverse family of parameters associated to a given similarity workload $({\mathbb X},d,{\mathbb U})$ and allowing to gauge the performance of various indexing schemes for similarity-based information retrieval.

Perhaps the best known single feature of a high-dimensional dataset $\mathbb U$ is ``the empty space paradox'': the average distance, $\E\e_{NN}(X)$, from a query point to the nearest neighbour in $\mathbb U$ grows along with the dimension of $\mathbb U$. Asymptotically, $\E\e_{NN}(X)$ approaches the average distance $\E [d(X,Y)]$ between two points of $\mathbb X$ (the {\em characteristic size} of dataset). 
This is a consequence of the fact that the distance functions $d_x(y) = d(x,y)$, $x\in{\mathbb X}$, concentrate near their median/mean values. The boxplot diagram (a) shows (normalized) pairwise distance distributions of $100$ uniformly sampled points in the cube ${\mathbb I}^d$ for various values of $d$. The lines in the middle of the boxes are median values, the lower and upper sides denote the first and third quartiles respectively, and the small circles mark outliers. For high $d$, even the outlier distance values approach the median.

\begin{figure}[ht]
\begin{center}
\scalebox{0.55}[0.55]{\includegraphics{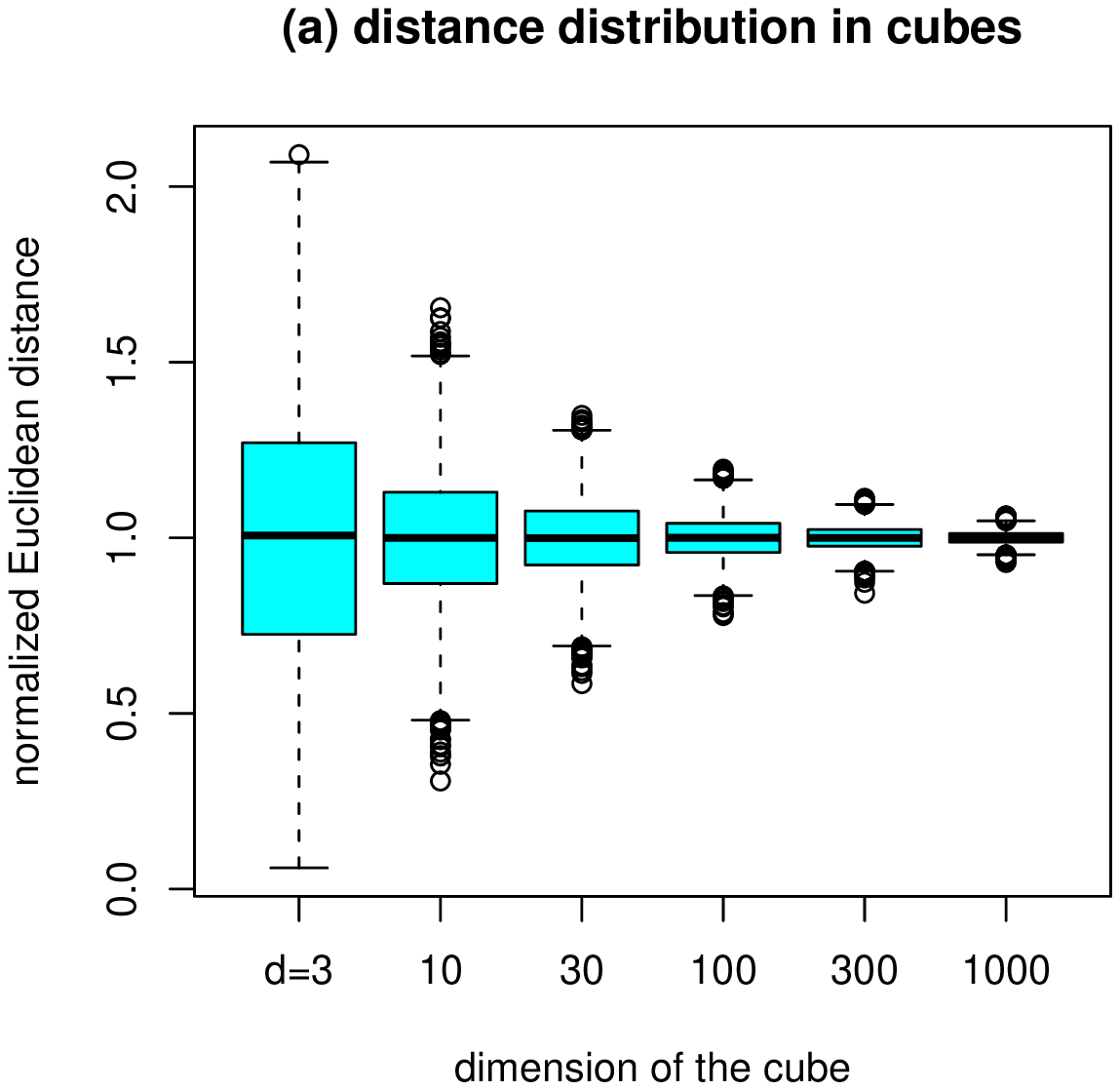}} 
\scalebox{0.55}[0.55]{\includegraphics{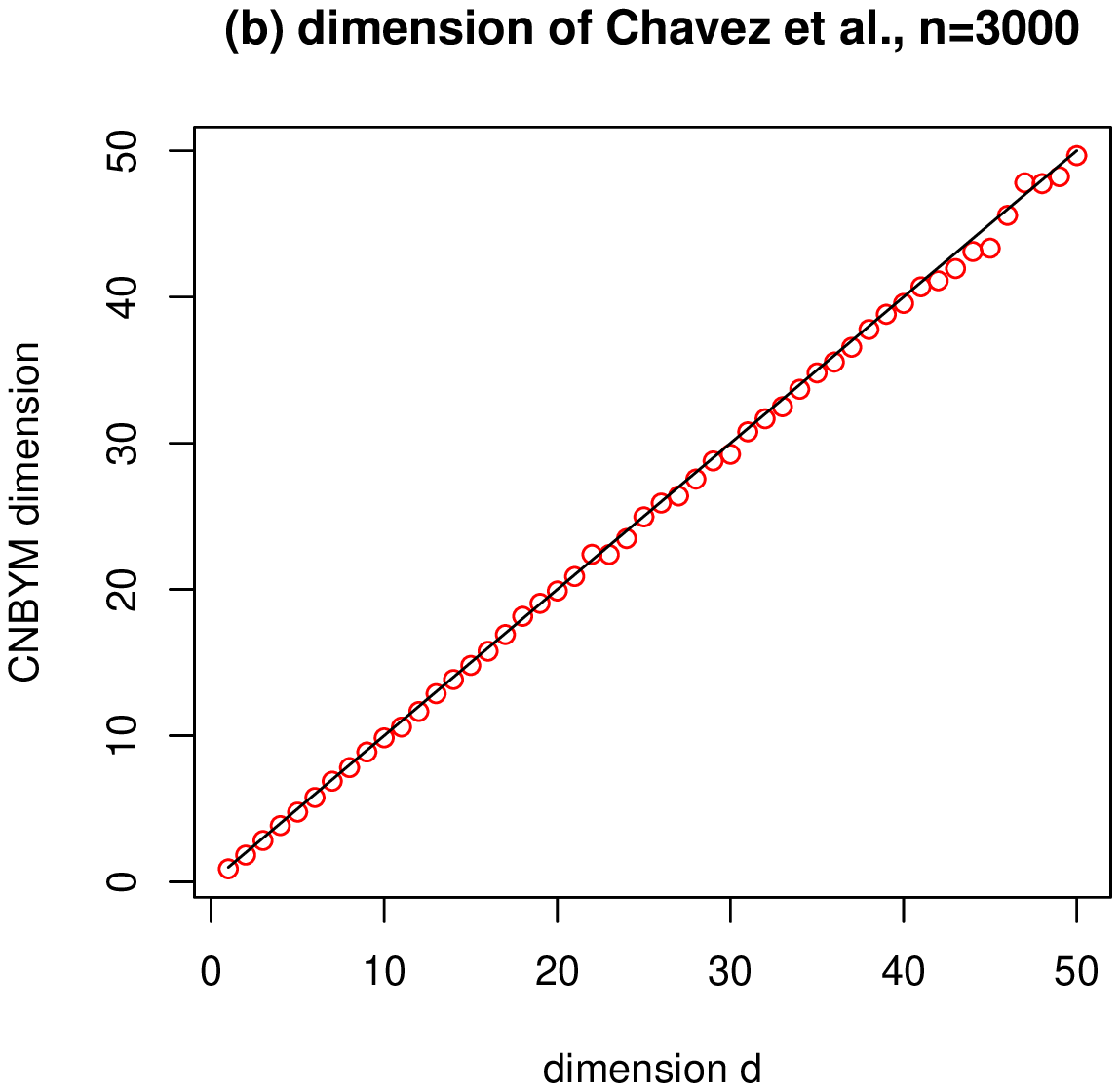}} 
\end{center}
\end{figure}  

Generally, the intrinsic dimensionality is 
inversely proportional to some dispersion parameter of data, usually normalized with regard to the characteristic size of $\mathbb U$. The possible choices of a dispersion parameter are many, and they reflect different aspects of the dataset and its indexability.

\smallskip

\noindent{\bf Illustration: the intrinsic dimensionality of Ch\'avez et al.}
The authors of \cite{CNBYM} have proposed and studied a parameter which we denote by the acronym of their last names: $\dim_{CNBYM}({\mathbb X}) =\E(d)^2/2{\mathrm{var}\,}(d)$. Here the distance $d$ is treated as a random variable on ${\mathbb X}\times{\mathbb X}$. The corresponding statistical parameter of a dataset $\mathbb U$ has a number of advantages: easy to compute, it makes a good statistical estimator for the dimension of the domain $\mathbb X$. The robustness of ``CNBYM dimension'' can be seen from diag. (b) where the parameter is calculated for independent random samples of $n=3,000$ points drawn from the gaussian distribution $\gamma^d$ on $\R^d$, $1\leq d\leq 50$.
\smallskip 

\noindent{\bf The curse of dimensionality.} 
When applied to high-dimensional datasets, many similarity-based information retrieval algorithms slow down to a point where they cannot outperform a simple sequential scan. Even though the phenomenon is well-known to data practitioners, curiously enough there is still no mathematical validation of it, and one of the major theoretical challenges in the field is the so-called {\em Curse of Dimensionality Conjecture} in the bare-bone setting where ${\mathbb X}=\{0,1\}^d$ is a Hamming cube \cite{indyk:04}. Rigorous lower bounds are only obtained in a small number of cases, some of which we will describe below.

\smallskip
\noindent{\bf The concentration phenomenon.}
The curse of dimensionality is closely linked to the {\em phenomenon of concentration of measure} (cf. e.g. \cite{gromov:99}, Ch. 3 $1/2$): on a typical geometric object $\mathbb X$ of high dimension not only the distance functions, but all the $1$-Lipschitz functions $f\colon {\mathbb X}\to\R$ concentrate near their median. (A function is {\em $1$-Lipschitz} if it does not increase distances: $\abs{fx-fy}\leq d(x,y)$.)
The {\em concentration function} $\alpha_{\mathbb X}(\e)$ of $\mathbb X$ gives an upper bound on the probability that the value of a $1$-Lipschitz function $f$ at an $x\in{\mathbb X}$ deviates from its median by more than $\e>0$. For instance, the concentration function of the $d$-dimensional Hamming cube $\{0,1\}^d$ with the normalized Hamming metric and uniform measure satisfies a Chernoff bound $\alpha(\e)\leq \exp(-2\e^2d)$, cf. diagram (c).
\begin{figure}[ht]
\begin{center}
\scalebox{0.55}[0.55]{\includegraphics{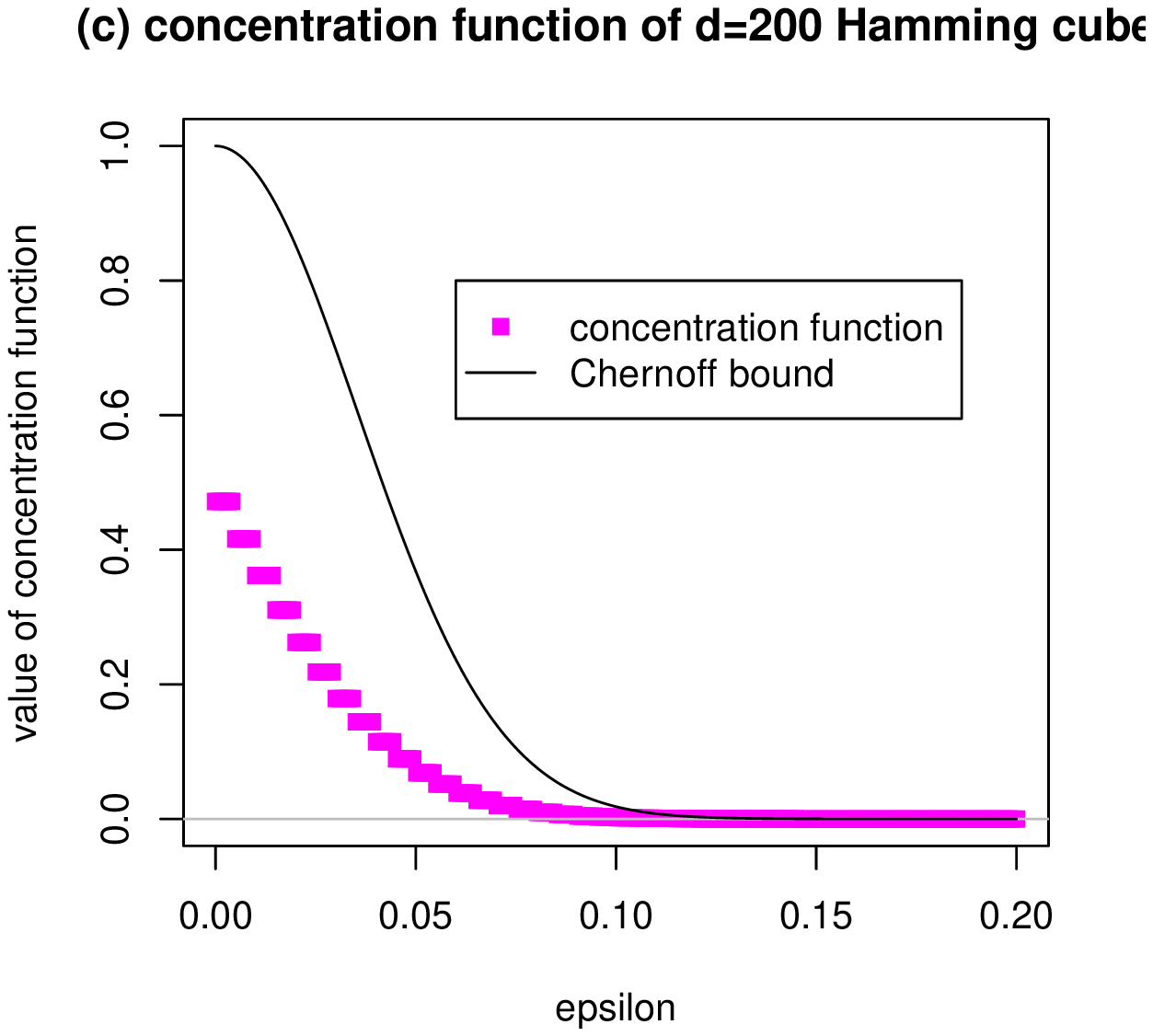}} 
\hskip 0.8cm
\raisebox{8ex}{\scalebox{0.28}[0.28]{\includegraphics{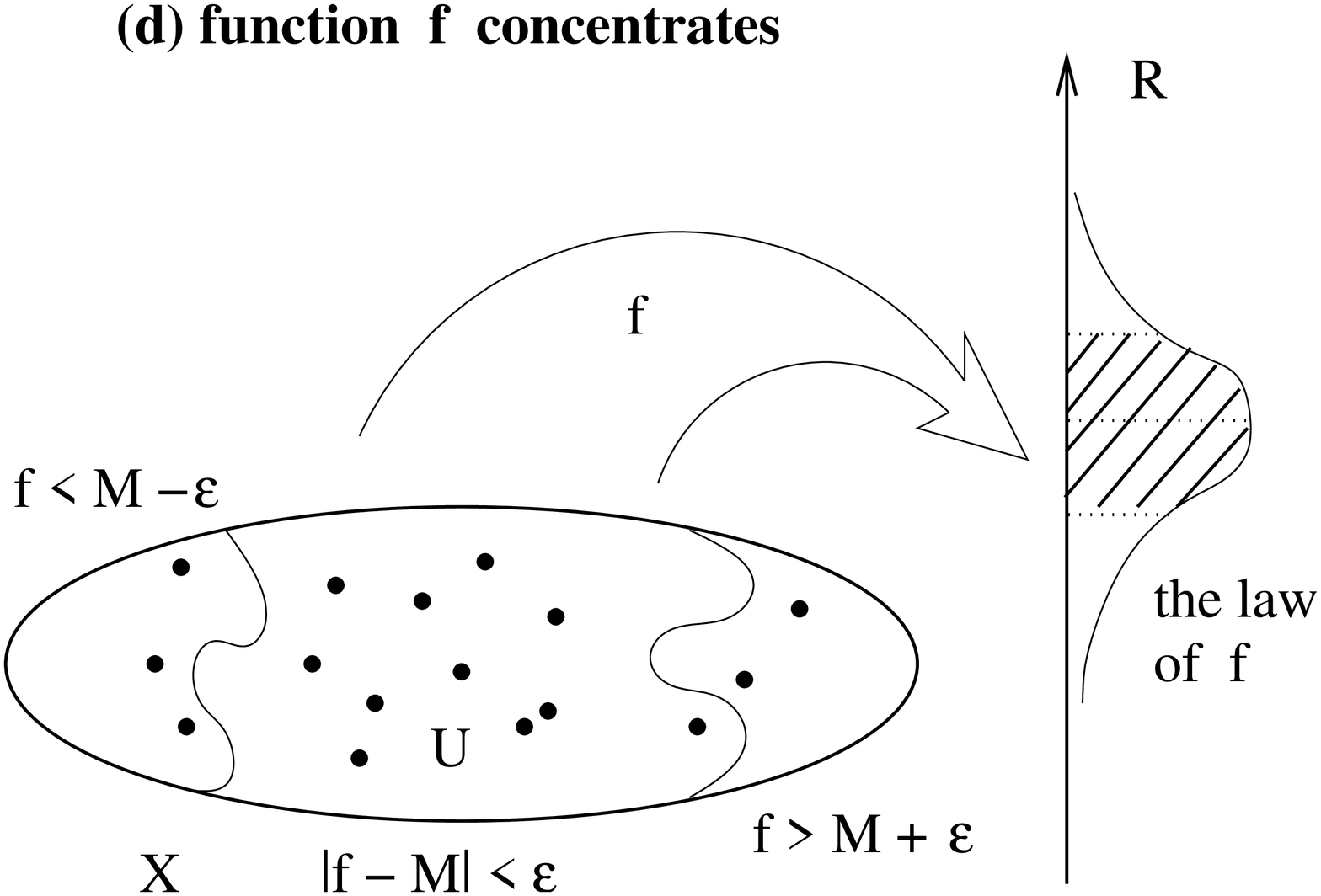}}}
\end{center}
\end{figure}  Similar gaussian bounds exist for Euclidean spheres, cubes, and a variety of other objects. The higher the dimension of $\mathbb X$, the more sharply $\alpha_{\mathbb X}(\e)$ drops off near zero. 

\smallskip
\noindent{\bf Degrading performance of indexing schemes.}
Many known access methods for {\em exact} similarity search (including pivot-based schemes, metric trees, etc.) have the following structure. An indexing scheme consists of a collection of  $1$-Lipschitz functions $\{f_a\colon a\in A\}$ on the domain $\mathbb X$ (possibly partially defined). Given a query point $x\in{\mathbb X}$ and the range value $\e>0$, calculations of the values $f_{a_i}(x)$, $i=1,2,\ldots,\ell$ are performed in a recursive way, where the function $f_{a_{i+1}}$ is chosen depending on the previous values $f_{a_1}(x),\ldots, f_{a_i}(x)$. If for some $i$ one has $\abs{f_{a_i}(y)-f_{a_i}(x)}>\e$, the datapoint $y$ cannot possibly belong to the $\e$-ball around $x$ and so is irrelevant. Datapoints $x_1,\ldots,x_k\in{\mathbb U}$ whose irrelevance cannot be certified in this way are returned, and each of them is checked against the condition $d(x,x_i)<\e$. 

If the domain $\mathbb X$ has high concentration dimension, then most values of each function $f_{a_i}$ concentrate near the median. Now denote $\mathscr F$  the class of all Lipschitz functions used to construct indexing schemes of a given type (e.g. the class of all distance functions for pivot schemes). Assume that $\mathscr F$ has a small capacity in the sense of statistical learning theory. (This condition is very natural and verified, e.g., by distance functions on the Euclidean spaces or Hamming cubes.)
Assume further that the points of $\mathbb U$ are sampled from $\mathbb X$ in an i.i.d. fashion. Then the values of each $f\in{\mathscr F}$ at datapoints can be shown to concentrate as well, and only a small proportion of points can be discarded. cf. diag. (d). This argument leads to lower bounds on the performance of pivot indexing schemes which are linear in the size of the dataset \cite{VolPest09}. Similar results can be deduced for metric trees. 

One is naturally led to {\em concentration dimension} of $\mathbb X$, definied by $\dim_{\alpha}({\mathbb X})=1/{\left[2\int_0^1 \alpha_{\mathbb X}(\e)~d\e\right]^2}$ and investigated in \cite{pestov0708}. Notice however that while $\dim_{\alpha}({\mathbb X})\geq \dim_{\alpha}({\mathbb U})$, the relationship between the two is not understood: the latter value (dimension of the empirical measure) need not be a good statistical estimator for the former.

\smallskip
\noindent
{\bf Concentration versus complexity.}
Heuristically, indexability of a workload $({\mathbb X},d,{\mathbb U})$ amounts to the existence of sufficiently many $1$-Lipschitz functions $f$ on $\mathbb U$ which at the same time {\em dissipate} (as opposed to {\em concentrate}) and have a low computational complexity. Thus, indexing means finding the right balance between concentration and complexity.

Here is an example of a ``success story'' where smallness of values of an intrinsic dimension guarantees the existence of an efficient access method. 

\smallskip
\noindent{\bf Assouad dimension.}
The {\em Assouad} ({\em doubling}) {\em dimension} of a metric space $\mathbb X$ is the minimum value $\rho\geq 0$ such that every set $A$ in $X$ can be covered by $2^{\rho}$ balls of half the diameter. (The {\em diameter}  of a set $A$ is the supremum of $d(x,y)$, $x,y\in A$.) To appreciate that $\dim_{Assouad}(X)$ is indeed an inverse quantity to some parameter of dispersion of data, notice that if $\mathbb X$ is equipped with a probability measure and $\dim_{CNBYM}({\mathbb X})$ is high, then the size of balls as a function of radius grows fast in the vicinity of $\E(d)$, and so $\dim_{Assouad}(X)$ has to be large.

As shown in \cite{KL}, a low Assouad dimension of a workload allows for the construction of a simple and efficient indexing scheme which is essentially a metric tree. The quantity $2^{\rho}$ bounds from above the number of leaf nodes (degree) of such a tree.

\smallskip
\noindent{\bf Concluding remarks and problems.} $\bullet$
A thorough survey of intrinsic dimensionality in the context of similarity search is \cite{clarkson}, though even this authoritative source is not comprehensive. 

\noindent
$\bullet$ It is safe to predict that the mainstream research direction in the field will be a discovery of further ``dimensionality parameters of positive type'', such as Assouad dimension. More and more real datasets that currently appear high-dimensional will turn out to have low, tractable intrinsic dimensions.

\noindent
$\bullet$
(Ciaccia, Pestov). Gromov's reconstruction theorem (\cite{gromov:99}, p. 120) says that a metric space with measure can be recovered from distribution laws of all $n$-point metric subspaces, for all $n$. (Compare with the CNBYM dimension, determined by the distribution law of $2$-point subspaces.) How much can one infer about indexability from the distribution law of $3$-point subspaces?

\noindent
$\bullet$ Notice that fractal-type dimensions (cf. \cite{ttf}) fit into a general paradigm of intrinsic dimensionality outlined by us at the beginning: they reflect the rate of growth of balls/boxes. So does a version of intrinsic dimension  proposed in \cite{lifshits}, the {\em disorder dimension,} whose aim is to quantify the situation where $\E(\e_{NN})$ is close in value to $\E(d)$.

\noindent
$\bullet$
We assumed above that queries follow the same underlying distribution as datapoints. Now suppose we have two underlying measures, $\mu$ modelling data and $\nu$ modelling query centers. Can one build a theory of concentration and intrisic dimension for such metric bi-measure spaces?

\end{document}